\documentclass{gretsi}
\small
\usepackage[english,french]{babel} 
\usepackage{times}			
  
\usepackage{graphicx}
\usepackage{mathtools}
\usepackage{amssymb}

\usepackage{algorithm,algpseudocode}
\usepackage{amsmath}
\usepackage{enumitem}
\usepackage{tikz}
\usepackage{pgfplots}
\usepackage{mathtools}

\usepackage[hidelinks]{hyperref}

\algnewcommand{\Inputs}[1]{%
	\State \textbf{Inputs:}
	\Statex \hspace*{\algorithmicindent}\parbox[t]{.8\linewidth}{\raggedright #1}
}
\algnewcommand{\Initialize}[1]{%
	\State \textbf{Initialize:}
	\Statex \hspace*{\algorithmicindent}\parbox[t]{.8\linewidth}{\raggedright #1}
}

\newcommand{\ma}[1]{\ensuremath{\mathsf{#1}}}
\renewcommand{\vec}[1]{\ensuremath{\mathbf{#1}}}

\newcommand{\ie}{\textit{i.e.}}
\newcommand{\eg}{\textit{e.g.}}

\newcommand\ygt[1]{\textcolor{black}{#1}}

\DeclareMathOperator*{\argmin}{\mathrm{argmin}}
\DeclareMathOperator*{\Var}{Var}

\DeclareMathOperator*{\diag}{diag}
\DeclareMathOperator*{\tr}{tr}

\usepackage{titlesec}

\title{Variance Reduction for Inverse Trace Estimation \\via Random Spanning Forests}

\author{\coord{Yusuf Yi\u git}{P\.ILAVCI}{}, 
        \coord{Pierre-Olivier}{AMBLARD}{},
        \coord{Simon}{BARTHELM\'E}{},
    \coord{Nicolas}{TREMBLAY}{}
    \thanks{This work was partly funded by
    	the French National Research Agency in the framework of the "Investissements d’avenir” program (ANR-15-IDEX-02), the LabEx PERSYVAL (ANR-11-LABX-0025-01), the ANR GraVa (ANR-18-CE40-0005),
    	the ANR GRANOLA  (ANR-21-CE48-0009),
    	the MIAI@Grenoble Alpes chairs ``LargeDATA at UGA" and ``Pollutants'' (ANR-19-P3IA-0003).}
    }

\address{\affil{}{Univ. Grenoble Alpes, CNRS, Grenoble INP, GIPSA-Lab}}

\email{firstname.lastname@gipsa-lab.fr}
\frenchabstract{La trace $\tr(q(\ma{L} + q\ma{I})^{-1})$, où $\ma{L}$ est une matrice symétrique à dominante diagonale, est une quantité d'intérêt dans certains problèmes d'apprentissage automatique. 
	Son calcul direct est difficile si la taille de la matrice est importante.
	Les méthodes de pointe comprennent l'estimateur de Hutchinson combiné à des solveurs itératifs, ainsi que l'estimateur basé sur les forêts aléatoires (un processus aléatoire sur les graphes).  
	Dans ce travail, nous révélons deux façons d'améliorer l'estimateur basé sur les forêts via des techniques bien connues de réduction de la variance, à savoir les variables de contrôle et l'échantillonnage stratifié. 
	La mise en œuvre de ces techniques est pratique et permet une réduction
  substantielle de la variance, donnant souvent des performances comparables à
  l'état de l'art ou meilleures.    
	}
\englishabstract{ The trace $\tr(q(\ma{L} + q\ma{I})^{-1})$, where $\ma{L}$ is a symmetric diagonally dominant matrix, is the quantity of interest in some machine learning problems. 
However, its direct computation is impractical if the matrix size is large.
State-of-the-art methods include Hutchinson's estimator combined with iterative solvers, as well as the estimator based on random spanning forests (a random process on graphs).  
In this work, we show two ways of improving the forest-based estimator via well-known variance reduction techniques, namely control variates and stratified sampling. 
Implementing these techniques is easy, and provides substantial
variance reduction, yielding comparable or better performance relative to
state-of-the-art algorithms.
}

\begin{document}
\maketitle
\section{Introduction}
Randomized methods are useful to approximate the trace of a matrix if the matrix is not explicitly known. These methods come into play in various problems~\cite{wu2016estimating} in which $\ma{A}\in\mathbb{R}^{n\times n}$ is typically a large matrix (\eg~$n\geq10^{6}$) and $\tr(f(\ma{A}))$ is the quantity of interest. 
In this work, we focus on calculating the trace of $f(\ma{L}) = q(\ma{L}+q\ma{I})^{-1}$ without taking the matrix inverse when $\ma{L}$ is a symmetric diagonally dominant (SDD) matrix~\ie~$\forall i, |\ma{L}_{i,i}| \geq \sum_{j\not=i}|\ma{L}_{i,j}|$.  
A natural use case of $\tr(q(\ma{L}+q\ma{I})^{-1})$ arises in graph Tikhonov
regularization problem~\cite{shuman2013emerging} where $\ma{L}$ is the graph Laplacian. 
In this problem, we are given a noisy signal over $n$ vertices $\vec{y}=[y_1,y_2,\ldots,y_n]^\top$ and we aim to recover the original signal $\vec{x}$ by solving the following problem: 
\begin{equation}
    \hat{\vec{x}} = \argmin_{\vec{z}\in\mathbb{R}^{n}}  q||\vec{y} - \vec{z}||^2_2 + \vec{z}^{\top}\ma{L}\vec{z}, \quad q>0
\end{equation}
where the hyper-parameter $q>0$ controls the regularization. 
The explicit solution $\hat{\vec{x}}$ reads $\ma{K}\vec{y}$ where $\ma{K}=q(\ma{L}+q\ma{I})^{-1}$. 
Notice that the recovery error,~\ie~$||\vec{x}-\hat{\vec{x}}||^2_2$, highly depends on $q$ and there are several methods to automatically choose the value of $q$ such that the solution $\hat{\vec{x}}$ approaches to $\vec{x}$. 
Many of them, such as generalized cross-validation, Akaike or Bayesian
information criteria, use $\tr(\ma{K})$ as a measure of the degrees of freedom
of the linear smoother $\ma{K}$~\cite{hastie2009elements}. 
\smallskip

\noindent \textbf{State-of-the-art.}  The standard estimator for $\tr(\ma{K})$ is due to Hutchinson~\cite{hutchinson1989stochastic}. Given $N$ samples of a Bernoulli random vector $\vec{a}\in\{1,-1\}^n$ with $\forall i$,  $\mathbb{P}(a_i = \pm1) = 1/2$, Hutchinson's estimator is defined as
$    h \coloneqq \frac{1}{N}\sum_{i=1}^N {\vec{a}^{(i)}}^\top\ma{K}\vec{a}^{(i)}.$
where $\vec{a}^{(i)}$s are samples of the random vector $\vec{a}$. The estimator $h$ is an unbiased estimator of $\tr(K)$. 
Note that one can change the law of $\vec{a}$ to any distribution satisfying
$\mathbb{E}[\vec{a}] = \vec{0}$ and $\Var(\vec{a}) = \ma{I}$ (\eg~Girard's
estimator~\cite{girard1987algorithme} arises when $\vec{a}$ is Gaussian with zero mean and unit variance). 

Computing $\ma{K}\vec{a}$ is expensive due the matrix inverse. 
Even leveraging the sparsity by using Cholesky decomposition has a time complexity $\mathcal{O}(n^3)$ in the worst case. 
For large $n$, this cost becomes prohibitive. The state-of-the-art that avoids this cubic cost consists of (preconditioned) conjugate gradient~\cite{shewchuk1994introduction}, algebraic multigrid~\cite{ruge1987algebraic}, polynomial approximations. 
They compute $\ma{K}\vec{a}$ with very small error, often much less than the
Monte Carlo error induced by Hutchinson's estimator, and they scale linearly with the number of edges $m$.
\smallskip

\noindent \textbf{RSF estimator.} In~\cite{barthelme:hal-02319194}, we proposed an alternative method to estimate $\tr(\ma{K})$ when $\ma{L}$ is a \ygt{SDD} matrix. 
This method is based on random spanning forests (RSF)~\cite{avena2018two}, a random process on graphs. We showed that the number of roots is an unbiased estimator for $\tr(\ma{K})$.
\smallskip

\noindent \textbf{Our contributions.} In this work, we improve the efficiency of the RSF-based estimator by well-known variance reduction (VR) techniques from the Monte Carlo literature.  
The main results of this paper are listed as follows:  
\begin{itemize}[label=\textbullet,leftmargin=0.35cm]
	\setlength\itemsep{-0.25em}
    \item We show two novel ways of applying VR techniques to the RSF-based estimator,
    \item The additional computations remain in the time complexity $\mathcal{O}(m)$ and come with practical implementations, 
    \item Empirical evidence on various graphs shows that the proposed methods
      perform at least as well as Hutchinson's estimator, while outperforming  it in many settings. 
\end{itemize}

\section{Background}
In this section, we introduce our notation and revisit some theoretical properties of RSFs. 
\smallskip

\noindent \textbf{Graph theory.} Consider an undirected, weighted graph $\mathcal{G}=(\mathcal{V},\mathcal{E},w)$ with $|\mathcal{V}| = n$ nodes and $|\mathcal{E}| = m$ edges. 
The weight function $w:\mathcal{V}\times\mathcal{V} \rightarrow \mathbb{R}_{\geq0}$ maps $\mathcal{E}$ to positive weights and for $(i,j)\not \in\mathcal{E}$, $w(i,j)$ equals to 0. 
The (weighted) adjacency matrix of a graph is the matrix $\ma{A} = [w(i,j)]_{i,j}\in\mathbb{R}^{n\times n}$.  
Degree of a node $i$ is $d_i=\sum_{j\mathcal{N}(i)}w(i,j)$ 
where $\mathcal{N}(i)$ is the neighborhood of $i$. We form the degree matrix as $\ma{D} = \diag(\vec{d})$.
Finally, the graph Laplacian $\ma{L}=\ma{D} - \ma{W}$ is a useful object with many applications in graph combinatorics, machine learning and graph signal processing~\cite{shuman2013emerging}.
\smallskip

\noindent \textbf{Random spanning forests.} 
A tree is a cycle-free subgraph of $\mathcal{G}$. It is a  \emph{spanning} tree if it reaches all vertices of $\mathcal{G}$.
A rooted tree is a directed tree whose edges are oriented towards a special node called a root. 
A rooted spanning forest, denoted by $\phi$, is a set of disjoint rooted trees on $\mathcal{G}$ whose union reaches all vertices.
Let us denote the set of all spanning forests by $\mathcal{F}$.
We define an \ygt{RSF} $\Phi_q$ as a random object that is defined over $\mathcal{F}$ and has the following distribution: 
\begin{equation}
	\mathbb{P}(\Phi_q = \phi) \propto q^{|\rho(\phi)|}\prod_{(i,j) \in \phi}w(i,j),  \quad 	q>0.	
	\label{eq:phiq}
\end{equation}
where $\rho(\phi)$ is the root set of $\phi$. 
Although $|\mathcal{F}|$ can be very large, a modified version of Wilson's algorithm~\cite{wilson1996generating} can be used to sample a forest~\cite{avena2018two}. 
The algorithm is based on loop-erased random walks on $\mathcal{G}$. 
Thus, the time complexity of the algorithm is reported as the expected number of steps until it terminates which is equal to $\tr(\ma{K}(\ma{I}+ \frac{1}{q}\ma{D}))\leq n+\frac{2m}{q}$~\cite{marchal2000loop}~\footnote{\ygt{$\tr(\ma{K}) + \frac{1}{q}\tr(\ma{KD}) \leq n + \frac{1}{q}\tr(\ma{KD})\leq n+\frac{1}{q}\tr(\ma{D}) = n+\frac{2m}{q}$.}}.

The random object $\Phi_q$ has fascinating theoretical properties that connect various concepts~\cite{avena2018two}.
An important example for this paper is $\mathbb{E}[|\rho(\Phi_q)|]$, which equals $\tr(\ma{K})$. 
Previously, we deployed $|\rho(\Phi_q)|$ as an unbiased estimate of $\tr(\ma{K})$. 
According to experiments performed on various graphs, this estimator is competitive and outperforms in some cases Girard's estimator in terms of the required time for reaching a certain precision. 
In this work, we improve the expected error of the RSF estimator by VR techniques for Monte Carlo estimators.

\section{Proposed Methods}
Two VR methods are applicable to the RSF estimator: The control variate (CV) technique and stratified sampling.\footnote{\ygt{We omit the main motivations behind these methods due to space limitations but we refer the reader to~\cite{kleijnen2010variance} for more details. }} 
Moreover, generalizing these methods to SDD matrices is straightforward~\cite{barthelme:hal-02319194}.   
Both methods use some additional information (\eg~a statistic with a known mean)
on the estimator to reduce variance. 
The difficulty in applying such methods is to find which additional statistic will be both fast to estimate and provide a substantial decrease in the variance.
This paper shows practical ways to adapt these techniques for the RSF estimator.

\subsection{Control Variates}
We give two RSF based unbiased estimators for $\ma{K}$ in~\cite{pilavci2021graph}. Both relies on the root relation $r_{\phi}:\mathcal{V}\rightarrow\rho(\Phi_q)$ which maps every node to its root in $\phi$. 
The first estimator is $\tilde{\ma{S}}\coloneqq[\mathbb{I}(r_{\Phi_q}(i) = j)]_{i,j}$ and verifies $\mathbb{E}[\tilde{\ma{S}}] = \ma{K}$ since $\mathbb{P}(r_{\Phi_q}(i) = j) = \ma{K}_{i,j}$. 
An improved version of this estimator with the CV method is~\cite{pilavci2021variance}: 
\begin{equation}
	\tilde{\ma{Z}}=\tilde{\ma{S}} -\alpha(\ma{K}^{-1}\tilde{\ma{S}}-\ma{I})
\end{equation}
Since $\mathbb{E}[\tilde{\ma{Z}}]=\ma{K}$, we find a unbiased trace estimator:
\begin{equation}
		\tilde{s} \coloneqq \tr(\ma{Z})  = |\rho(\Phi_q)| - \alpha \tilde{c},
\end{equation} 
where
\[
\tilde{c}= \left(n - |\rho(\Phi_q)|  -\frac{1}{q}\sum_{\substack{{i\in\rho(\Phi_q)}\\{j\in\mathcal{N}(i)}}}w(i,j)\mathbb{I}(r_{\Phi_q}(j)\not=i)\right).
\]
The random variable $\tilde{c}$ is called the ``control variate'', and its mean is $n$.  
To calculate $\tilde{c}$, one only needs to count the neighbors of each root $i$ that are not rooted in $i$.       
For $|\rho(\Phi_q)|\ll n$, the computational cost remains negligible, whereas, in the worst case, it might require traversing every edge of the graph. 
One can also adapt these calculations for the second estimator in~\cite{pilavci2021graph}. To do so, let us recall this estimator in matrix form; the trees of $\Phi_q$ depict a random partition $\mathcal{P} = \{\mathcal{V}_1,\mathcal{V}_2,\ldots,\mathcal{V}_{|\rho(\Phi_q)|}\}$ over $\mathcal{V}=\cup_{i=1}^{|\rho(\Phi_q)|}\mathcal{V}_{i}$. 
We enumerate these components from 1 to $|\rho(\Phi_q)|$ and consider a mapping
$t$ from each vertex $i$ to the number of the component that $i$ belongs to.
Then, the second estimator takes the form:
$\bar{\ma{S}} = \left[\frac{\mathbb{I}(i\in\mathcal{V}_{t(j)})}{|\mathcal{V}_{t(j)}|}\right]_{i,j}$. 
So, one has: 
\begin{equation}
	\bar{s} \coloneqq |\rho(\Phi_q)| -\alpha\bar{c},
\end{equation}
where 
$
\bar{c}=n - |\rho(\Phi_q)| -\frac{1}{q} \sum_{\substack{{i\in\mathcal{V}}\\{j\in \mathcal{N}(i)}}}\bar{\ma{S}}_{i,i}w(i,j)\mathbb{I}(r_{\Phi_q}(j)\not=i)$.
In this case, the control variate requires keeping track of partition sizes and
neighbors at partition boundaries. 
While the former can be done in $\mathcal{O}(n)$, the latter requires traversing all edges. 
However, it provides more variance reduction than the previous option
(See Prop. 1 and 2 in~\cite{pilavci2021graph}).
\smallskip

\noindent \textbf{How to choose $\alpha$}. As can be deduced from Prop. 2 in ~\cite{pilavci2021variance}, a safe value of $\alpha$,~\ie~a value that guarantees variance reduction, is $\frac{2q}{q+d_{max}}$ where $d_{max}$ is the maximum degree in $\mathcal{G}$. 
We also observe that $\frac{q}{q+d_{avg}}$ is usually a good estimate of $\alpha^{\star}$ where $d_{avg}$ is the average degree in $\mathcal{G}$. 
\subsection{Stratified Sampling}
Stratification reduces the Monte Carlo error by dividing the sample space into sub-parts, each called a stratum, based on another random variable. 
Stratified sampling can substantially decrease approximation error when applicable.
In the following, we give a way of applying stratification to the RSF-based trace estimator.
\smallskip

\noindent \textbf{Stratification for the RSF estimator.} Consider the root set that are sampled at the first visit of random walks in Wilson's algorithm.  
Let us denote them by $\rho'(\Phi_q)$ and define a random variable $R_i\coloneqq\mathbb{I}(i\in\rho'(\Phi_q))$ where $\mathbb{I}$ is the indicator function.  
Notice that each $R_i$ is an independent Bernoulli variable with $\mathbb{P}(R_i=1) = \frac{q}{q+d_i}$. Building on this,
we propose to use the cardinality $|\rho'(\Phi_q)|=\sum_{i\in\mathcal{V}}R_i\in\{0,1,\ldots,n\}$ to apply stratification on the RSF estimator as follows; i/ take disjoint $K$-fold strata $C_1,\dots,C_K$ verifying $\bigcup_{i=1}^K C_i = \{0,1,\dots,n\}$, ii/ get $N_i$ samples of $\Phi_q\Big||\rho'(\Phi_q)|\in C_i$ for each stratum $C_i$, iii/ compute the following weighted sum: 
\begin{equation}
	s_{st} \coloneqq \sum_{i=1}^K\frac{1}{N_i}\left(\sum_{\substack{j=1\\|\rho'(\phi^{(j)})|\in C_i }}^{N_i} |\rho(\phi^{(j)})|  \right)\mathbb{P}(|\rho'(\Phi_q)|\in C_i).
\end{equation}
For $N=\sum_{i=1}^KN_i$ samples, $s_{st}$ gives an unbiased estimation of $\tr(\ma{K})$ due to the law of conditional expectation. 
Moreover, given a fixed $N$, certain settings of $N_i$'s provide lower theoretical
variance,~\eg~$N_i=N\mathbb{P}(|\rho'(\Phi_q)|\in
C_i)$~\cite{kleijnen2010variance}. 
\smallskip 

\noindent \textbf{Implementation.} We address two issues in implementing stratified sampling. 
The first one is the calculation of the probabilities $\mathbb{P}(|\rho'(\Phi_q)|\in C_i)$. 
We approximate the distribution of $|\rho'(\Phi_q)|$ by a normal distribution with a mean $\mu = \sum_{i\in\mathcal{V}} \frac{q}{q+d_i}$ and a variance $\sigma^2=\sum_{i\in\mathcal{V}} \frac{qd_i}{(q+d_i)^2}$ to avoid expensive calculations of the exact methods~\cite{hong2013computing}. 
The second is to sample the random variable $|\rho(\Phi_q)| \Big| |\rho'(\Phi_q)|\in C_i$. 
Given a set $\mathcal{X}\subseteq\mathcal{V}$ verifying $|\mathcal{X}|\in C_i$, we can easily adapt Wilson's algorithm for sampling $\Phi_q|\rho'(\Phi_q) = \mathcal{X}$ with two modifications; i/ we pass $\mathcal{X}$ as the initial root set of $\Phi_q$, ii/ we prevent any node $i\not\in\mathcal{X}$ being a root at the first visit of walks in Wilson's algorithm. 
For the generation of the fixed set $\mathcal{X}$, we use rejection
sampling~\cite{bishop2006pattern} which is fast if $\forall i$, $\mathbb{P}(|\rho'(\Phi_q)|\in C_i) \gg 0$. 

\section{Experiments}
\begin{figure*}[t]
	\centering
	\includegraphics[width=17cm,height=4.4cm]{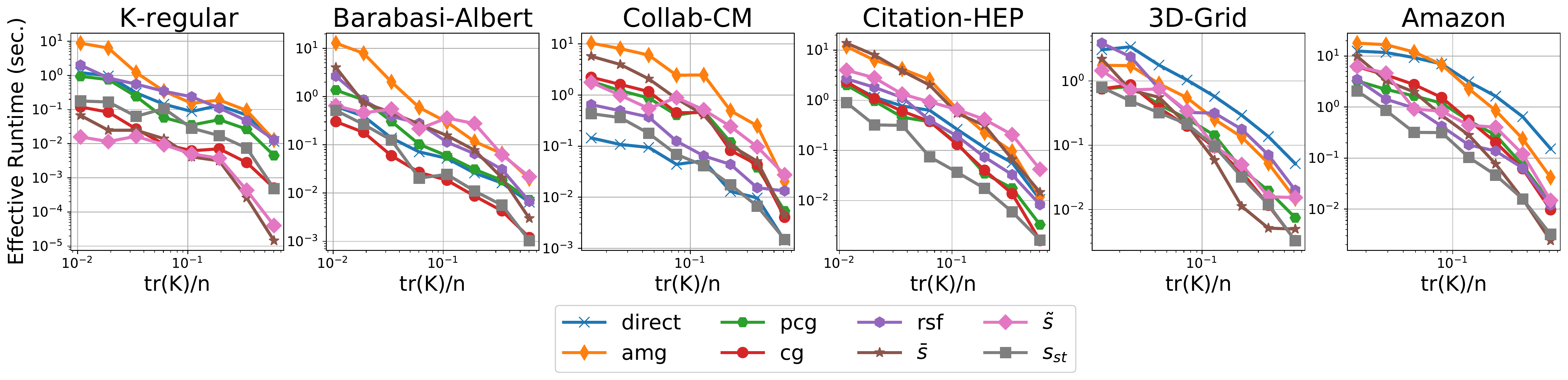}
	\caption{Effective Runtime vs $\tr(\ma{K})/ n$. }
	\label{fig:eff_time}
\end{figure*}
We empirically compare the proposed methods to Hutchinson's estimator over various graphs by following a similar procedure to~\cite{barthelme:hal-02319194}. 
Notice that all estimators here are Monte Carlo. 
Therefore, the asymptotic relation between the variance of a Monte Carlo estimator over a single sample $\sigma_1$, and $N$ samples $\sigma_N$, ~\ie~$\sigma_N\approx\frac{\sigma_1}{N^{1/2}}$, applies to all the estimators in the comparison
\footnote{This holds for the stratified sampling with $N_i=N\mathbb{P}(|\rho'(\Phi_q)|\in
	C_i)$ for all $i$. However, it is not necessarily true for other choices of $N_i$'s.}.  
We leverage this fact to compare the effective runtimes of all methods~\ie~the time needed to reach a fixed relative error $\epsilon$. 
First, we run all methods with $N=100$. This gives us the average runtime for the computation per sample and the sample variance $\hat{\sigma}^2_N$. Then, we approximate $\hat{\sigma}_1 = \sqrt{N}\hat{\sigma}_N$ for each method. 
By using this approximation, we solve $
\epsilon = \frac{\hat{\sigma}_1}{\tr(K)\sqrt{k}}$
for $\epsilon=0.002$ to calculate the number of iterations $k$ needed for reaching $\epsilon$ error. 
Finally, we calculate the effective runtime per method by multiplying $k$ by the average time for generating a single sample. 

In Hutchinson's estimator, we compute $\ma{K}\vec{a}$ using;
Algebraic Multigrid
(AMG)~\footnote{\url{https://github.com/JuliaLinearAlgebra/AlgebraicMultigrid.jl}},
Conjugate Gradient
(CG)~\footnote{\url{https://docs.juliahub.com/KrylovMethods}}, CG with AMG
preconditioning, and finally sparse Cholesky decomposition using CHOLMOD~\cite{chen2008algorithm}. 
Here, the CG methods benefit from block implementations ~\cite{o1980block}. 
We compare these with our proposed methods over various graphs. 
For $\tilde{s}$ and $\bar{s}$, we set $\alpha=\frac{q}{q+d_{avg}}$. 
In stratified sampling, we divide the sample space into 5 strata $C_1,\dots,C_5$ verifying $\mathbb{P}(|\rho'(\Phi_q)|\in C_k)\approx0.2$ for all $k={1,\dots,5}$. 
We set $N_k=N\mathbb{P}(|\rho'(\Phi_q)|\in C_k)$ per stratum $k$.
The graphs that we use in these experiments are:
\begin{itemize}[label=\textbullet,leftmargin=0.35cm]
	\setlength\itemsep{-0.25em}
	\item \textbf{Barabasi-Albert}: A random graph generated by Barabasi-Albert model ($k=10$) with $n=10^4$ and $m=99900$, 
	\item \textbf{K-random regular}: A random regular graph with $n=10^4$ and $m=10^5$ ($k=20$),
	\item \textbf{Collab-CM}: A collaboration network of $n=21363$ authors in Arxiv on condense matter physics with $m=91342$ links,
	\item \textbf{Citation-HEP}: A citation network of $n=34401$ in Arxiv on high energy physics with $m=420828$ links.   
	\item \textbf{3D Grid}: 3-dimensional grid with $n=50^3=125000$ nodes and $m=375000$ edges. 
	\item \textbf{Amazon}: A real-life network over $n=262111$ products in Amazon with $m=899792$. 
	A link between two products indicates that the same client purchases these two products.\footnote{The real-life data sets can be found in~\url{https://snap.stanford.edu/data/}}  
\end{itemize}
We choose 8 logarithmically spaced values of $q$ such that the ratio $\tr(\ma{K})/ n$ takes values up to $65\%$.
All experiments are implemented in Julia and run in a single thread of a laptop. 

\ygt{Fig.~\ref{fig:eff_time} summarizes} the results. For relatively small and sparse graphs, such as Collab-CM, the direct method gives the best performance, closely followed by the RSF methods. However, the approximate ones beat the direct method when the graphs become larger or denser. 
In these cases, the proposed methods give either the best or a comparable performance with the other state-of-the-art methods. 
A comparison between the regular and highly irregular graphs,~\eg~K-regular vs Barabasi-Albert, shows that the CV estimators $\tilde{s}$ and $\bar{s}$ gives small expected error in regular cases. 
\ygt{This is an expected result since $\tilde{c}$ and $\bar{c}$  have lower variances on regular graphs as they are summations over \ygt{the neighbors of the roots}.} 
In \ygt{irregular graphs}, the stratified sampling estimator often outperforms state-of-the-art. 
\vspace{-0.3cm}
\section{Conclusion}
The rich theoretical properties of RSFs give us several ways to improve the RSF trace estimator. 
In the future, we plan to develop estimators for other Laplacian based quantities, such as the elements of $\ma{K}$, or the effective resistances.    
We also note that we use relatively naive implementations for the stratified
sampling method,~\eg~ the normal approximation for the Poisson-Binomial
distribution can be improved by using \eg~ Cornish-Fisher or saddlepoint
approximations~\cite{dasgupta2008asymptotic}.

\vspace{-0.4cm}
\bibliographystyle{abbrv}
\bibliography{Refs}

\begin{thebibliography}{10}

\bibitem{avena2018two}
L.~Avena and A.~Gaudilli{\`e}re.
\newblock Two applications of random spanning forests.
\newblock {\em Journal of Theoretical Probability}, 31(4):1975--2004, 2018.

\bibitem{barthelme:hal-02319194}
S.~Barthelme, N.~Tremblay, A.~Gaudilliere, L.~Avena, and P.-O. Amblard.
\newblock {Estimating the inverse trace using random forests on graphs}.
\newblock In {\em {GRETSI 2019 - XXVII{\`e}me Colloque francophone de
  traitement du signal et des images}}, Lille, France, Aug. 2019.

\bibitem{bishop2006pattern}
C.~M. Bishop and N.~M. Nasrabadi.
\newblock {\em Pattern recognition and machine learning}, volume~4.
\newblock Springer, 2006.

\bibitem{chen2008algorithm}
Y.~Chen, T.~A. Davis, W.~W. Hager, and S.~Rajamanickam.
\newblock Algorithm 887: Cholmod, supernodal sparse cholesky factorization and
  update/downdate.
\newblock {\em ACM Transactions on Mathematical Software (TOMS)}, 35(3):1--14,
  2008.

\bibitem{dasgupta2008asymptotic}
A.~DasGupta.
\newblock {\em Asymptotic theory of statistics and probability}, volume 180.
\newblock Springer, 2008.

\bibitem{girard1987algorithme}
D.~Girard.
\newblock Un algorithme simple et rapide pour la validation crois{\'e}e
  g{\'e}n{\'e}ralis{\'e}e sur des probl{\`e}mes de grande taille.
\newblock Technical report, 1987.

\bibitem{hastie2009elements}
T.~Hastie, R.~Tibshirani, J.~H. Friedman, and J.~H. Friedman.
\newblock {\em The elements of statistical learning: data mining, inference,
  and prediction}, volume~2.
\newblock Springer, 2009.

\bibitem{hong2013computing}
Y.~Hong.
\newblock On computing the distribution function for the poisson binomial
  distribution.
\newblock {\em Computational Statistics \& Data Analysis}, 59:41--51, 2013.

\bibitem{hutchinson1989stochastic}
M.~F. Hutchinson.
\newblock A stochastic estimator of the trace of the influence matrix for
  laplacian smoothing splines.
\newblock {\em Communications in Statistics-Simulation and Computation},
  18(3):1059--1076, 1989.

\bibitem{kleijnen2010variance}
J.~P. Kleijnen, A.~Ridder, and R.~Rubinstein.
\newblock Variance reduction techniques in monte carlo methods.
\newblock 2010.

\bibitem{marchal2000loop}
P.~Marchal.
\newblock Loop-erased random walks, spanning trees and hamiltonian cycles.
\newblock {\em Electronic Communications in Probability}, 5:39--50, 2000.

\bibitem{o1980block}
D.~P. O'Leary.
\newblock The block conjugate gradient algorithm and related methods.
\newblock {\em Linear algebra and its applications}, 29:293--322, 1980.

\bibitem{pilavci2021variance}
Y.~Pilavc{\i}, P.-O. Amblard, S.~Barthelm{\'e}, and N.~Tremblay.
\newblock Variance reduction in stochastic methods for large-scale regularised
  least-squares problems.
\newblock {\em arXiv preprint arXiv:2110.07894}, 2021.

\bibitem{pilavci2021graph}
Y.~Y. Pilavc{\i}, P.-O. Amblard, S.~Barthelme, and N.~Tremblay.
\newblock Graph tikhonov regularization and interpolation via random spanning
  forests.
\newblock {\em IEEE transactions on Signal and Information Processing over
  Networks}, 7:359--374, 2021.

\bibitem{ruge1987algebraic}
J.~W. Ruge and K.~St{\"u}ben.
\newblock Algebraic multigrid.
\newblock In {\em Multigrid methods}, pages 73--130. SIAM, 1987.

\bibitem{shewchuk1994introduction}
J.~R. Shewchuk et~al.
\newblock An introduction to the conjugate gradient method without the
  agonizing pain, 1994.

\bibitem{shuman2013emerging}
D.~I. Shuman, S.~K. Narang, P.~Frossard, A.~Ortega, and P.~Vandergheynst.
\newblock The emerging field of signal processing on graphs: Extending
  high-dimensional data analysis to networks and other irregular domains.
\newblock {\em IEEE signal processing magazine}, 30(3):83--98, 2013.

\bibitem{wilson1996generating}
D.~B. Wilson.
\newblock Generating random spanning trees more quickly than the cover time.
\newblock In {\em Proceedings of the twenty-eighth annual ACM symposium on
  Theory of computing}, pages 296--303, 1996.

\bibitem{wu2016estimating}
L.~Wu, J.~Laeuchli, V.~Kalantzis, A.~Stathopoulos, and E.~Gallopoulos.
\newblock Estimating the trace of the matrix inverse by interpolating from the
  diagonal of an approximate inverse.
\newblock {\em Journal of Computational Physics}, 326:828--844, 2016.

\end{thebibliography}

\end{document}